\documentclass[12pt]{article}
\usepackage{amsmath,amsfonts,amssymb,makeidx,ifthen}
\usepackage[dvipdfmx]{graphicx}
\makeatletter
\@addtoreset{equation}{section}
\makeatother

\makeatletter
\long\def\@makecaption#1#2{{\small
\advance\leftskip1cm
\advance\rightskip1cm
\vskip\abovecaptionskip
\sbox\@tempboxa{#1: #2}%
\ifdim \wd\@tempboxa >\hsize
 #1: #2\par
\else
\global \@minipagefalse
\hb@xt@\hsize{\hfil\box\@tempboxa\hfil}%
\fi
\vskip\belowcaptionskip}}
\makeatother
\def\eq#1\en{\begin{equation}#1\end{equation}}  
\def\eqa#1\ena{\begin{align}#1\end{align}}
\def\eqg#1\eng{\begin{gather}#1\end{gather}}
\newcommand{\lb}[1]{\label{e:#1}}
\newcommand{\rlb}[1]{\eqref{e:#1}} 
\newcommand{\nl}{\notag\\}

\newcommand{\abs}[1]{\left|#1\right|}

\newcommand{\norms}[1]{\Vert#1\Vert}

\newcommand{\rbk}[1]{\left(#1\right)}

\newcommand{\bkt}[1]{\left\langle#1\right\rangle}
\newcommand{\sbkt}[1]{\langle#1\rangle}
\newcommand{\bbkt}[1]{\bigl\langle#1\bigr\rangle}

\newcommand{\sumtwo}[2]%
{\mathop{\sum_{#1}}_{#2}}
\newcommand{\sumtwoone}[3]%
{\mathop{\sum_{#1}^{#2}}_{#3}}
\newcommand{\sumthree}[3]%
{\mathop{\mathop{\sum_{#1}}_{#2}}_{#3}}
\newcommand{\sumfour}[4]%
{\mathop{\mathop{\mathop{\sum_{#1}}_{#2}}_{#3}}_{#4}} 
\newcommand{\prodtwo}[2]%
{\mathop{\prod_{#1}}_{#2}}
\newcommand{\mintwo}[2]%
{\mathop{\min_{#1}}_{#2}}
\newcommand{\maxtwo}[2]%
{\mathop{\max_{#1}}_{#2}}
\newcommand{\maxthree}[3]%
{\mathop{\mathop{\max_{#1}}_{#2}}_{#3}}
\newcommand{\limtwo}[2]%
{\mathop{\lim_{#1}}_{#2}}
\newcommand{\suptwo}[2]%
{\mathop{\sup_{#1}}_{#2}}
\newcommand{\supthree}[3]%
{\mathop{\mathop{\sup_{#1}}_{#2}}_{#3}}
\newcommand{\supfour}[4]%
{\mathop{\mathop{\mathop{\sup_{#1}}_{#2}}_{#3}}_{#4}} 
\newcommand{\inftwo}[2]%
{\mathop{\inf_{#1}}_{#2}}
\newcommand{\infthree}[3]%
{\mathop{\mathop{\inf_{#1}}_{#2}}_{#3}}
\newcommand{\inffour}[4]%
{\mathop{\mathop{\mathop{\inf_{#1}}_{#2}}_{#3}}_{#4}} 

\newcommand\calH{{\cal H}}

\newcommand{\tiu}{\tilde{u}}

\newcommand{\tiU}{\tilde{U}}

\newcommand{\tisigma}{\tilde{\sigma}}

\newcommand{\tiphi}{\tilde{\varphi}}

\newcommand{\bbR}{\mathbb{R}}
\newcommand{\bbZ}{\mathbb{Z}}
\newcommand{\ep}{\varepsilon}
\newcommand{\up}{\uparrow}

\newcommand{\Di}{\mathit{\Delta}}
\newcommand{\qedm}{\rule{1.5mm}{3mm}}

\newcommand{\Du}{\Di u}

\newcommand{\bra}[1]{\langle#1|}
\newcommand{\ket}[1]{|#1\rangle}

\newcommand{\kph}{\ket{\varphi}}
\newcommand{\bph}{\bra{\varphi}}

\newcommand{\kj}{\ket{j}}
\newcommand{\jAj}{\bra{j}\hA\ket{j}}

\newcommand{\hHN}{\hat{H}^{(N)}}
\newcommand{\hH}{\hat{H}}
\newcommand{\hHp}{\hat{H}^{\rm per}}
\newcommand{\EN}{E^{(N)}}
\newcommand{\hA}{\hat{A}}
\newcommand{\hB}{\hat{B}}
\newcommand{\hC}{\hat{C}}
\newcommand{\hX}{\hat{X}}
\newcommand{\hY}{\hat{Y}}

\newcommand{\bktcan}[1]{\sbkt{#1}^{\rm can}_{N,\beta_0}}
\newcommand{\GN}{\Gamma_N}
\newcommand{\ON}{\Omega_N}

\newcommand{\dist}{\operatorname{dist}}
\newcommand{\supp}{\operatorname{supp}}
\newcommand{\bz}{\beta_0}
\newcommand{\kT}{k_{\rm B}T_0}
\newcommand{\umin}{u_\mathrm{min}}
\newcommand{\umax}{u_\mathrm{max}}
\newcommand{\umid}{u_\mathrm{mid}}
\newcommand{\limN}{\lim_{N\up\infty}}
\newcommand{\limn}{\lim_{n\up\infty}}
\newcommand{\Tr}{\operatorname{Tr}}

\begin{document}

\noindent
{\bf
\Large On the local equivalence between the  canonical and the microcanonical ensembles for quantum spin systems}
\par\bigskip

\noindent
Hal Tasaki\footnote{
Department of Physics, Gakushuin University, Mejiro, Toshima-ku, 
Tokyo 171-8588, Japan
}
\begin{quotation}
We study a quantum spin system on the $d$-dimensional hypercubic lattice $\Lambda$ with $N=L^d$ sites with periodic boundary conditions.
We take an arbitrary translation invariant short-ranged Hamiltonian.
For this system, we consider both the canonical ensemble with inverse temperature $\bz$ and the microcanonical ensemble with the corresponding energy $U_N(\bz)$.
For an arbitrary self-adjoint operator $\hA$ whose support is contained in a hypercubic block $B$ inside $\Lambda$, we prove that the expectation values of $\hA$ with respect to these two ensembles are close to each other for large $N$ provided that  $\bz$ is sufficiently small and the number of sites in $B$ is $o(N^{1/2})$. 
This establishes the equivalence of ensembles on the level of local states in a large but finite system.
The result is essentially that of Brandao and Cramer (here restricted to the case of the canonical and the microcanonical ensembles), but we prove improved estimates in an elementary manner.
We also review and prove standard results on the thermodynamic limits of thermodynamic functions and the equivalence of ensembles in terms of thermodynamic functions.
The present paper assumes only elementary knowledge on quantum statistical mechanics and quantum spin systems.
\end{quotation}

\tableofcontents

\bigskip

\section{Introduction}
In statistical mechanics the ``equivalence of ensembles'' stands for a statement that different equilibrium ensembles, e.g., the canonical and the microcanonical ensembles, for a given macroscopic system lead to the same predictions.\footnote{
This should not be confused with the (related) statement that, when the whole system is described by the microcanonical ensemble, its small part is approximately described by the canonical ensemble.
}
As a celebrated example, it is known for a  very general class of models that the free energy density and the entropy density (in their infinite volume limits) are related via the Legendre transformation \cite{Ruelle,Lima71}.
See \rlb{Legendre2} and \rlb{Legendre} below.
This is true even at phase transition points or in phase coexistence regions.
This relation illustrates a deep connection between statistical mechanics and thermodynamics, and touches the essence of equilibrium physics.\footnote{
We believe that there should be undergraduate textbooks which contain a  proof of the equivalence, and there indeed is \cite{Tasaki}.
}

One can also discuss the equivalence on the level of expectation values (or states), i.e., one considers the canonical and the microcanonical ensembles of a given macroscopic system and asks if the respective expectation values $\sbkt{\hA}^{\rm can}$ and $\sbkt{\hA}^{\rm mc}$ of an observable $\hA$ coincide (or are close).\footnote{
Here the inverse temperature for the canonical ensemble and the energy density for the microcanonical ensemble should be precisely tuned according to the standard prescription of the equivalence of ensembles.
}
It was indeed shown that, in the limit where the size of the whole system becomes infinite, the expectation values of any local operators precisely coincide, under the essential assumption that the system has a unique infinite volume equilibrium state.
This was proved for classical lattice systems by Martin-L\"of \cite{MartinLof}, for classical particle systems by Gerogii \cite{Georgii}, and for quantum spin systems by Lima \cite{Lima72} and Mueller, Adlam, Masanes, and Wiebe \cite{MuellerAdlamMasanesWiebe}.
It is likely that the assumption of unique equilibrium state is essential rather than technical.
The equivalence of ensembles on the level of states may not be as straightforward as that on the level of thermodynamic functions.

More recently Brandao and Cramer \cite{BrandaoCramer} formulated and solved a finite volume version of the problem.
They considered a large but finite system with $N$ sites, and showed that the expectation values $\sbkt{\hA}^{\rm can}$ and $\sbkt{\hA}^{\rm mc}$ are close for operator $\hA$ whose support is contained in a smaller block with $o(N^{1/(d+1)})$ sites.
Again the proof is based on an essential assumption that any truncated two-point correlation function exhibit exponential decay.
In fact Brandao and Cramer solved much more general problem of comparing different states on a large but finite lattice, and above mentioned equivalence of the canonical and the microcanonical ensembles is one application.

In the present paper, we shall rederive the result of Brandao and Cramer, restricting ourselves only to the equivalence of the canonical and the microcanonical ensembles.
By concentrating on translation invariant quantum spin systems with periodic boundary conditions and using ``fine-tuned'' range for the microcanonical ensemble, we get better bounds than those in  \cite{BrandaoCramer}.
In our case the small block may contain $o(N^{1/2})$ sites, and the energy width of the microcanonical ensemble can be anything not less than an arbitrary fixed constant.

The proof of Brandao and Cramer is based on advanced techniques of quantum information.
The main tool is the Berry-Essen theorem for quantum lattice systems, and a smoothed quantum relative entropy is used to compare different quantum states.
It is not easy to completely understand the proof unless one is familiar to modern development in quantum information theory.

Our proof, on the other hand, makes only use of standard techniques of statistical mechanics, and is indeed quite elementary.
Our basic strategy is to combine the well-known technique in the proof of the equivalence of ensembles with an elementary large-deviation type estimate \rlb{LD} to prove the key estimate \rlb{ZOe}.
In the proof we make full use of detailed information about the free energy (the Massieu function) obtained from the cluster expansion method, as well as the decay properties of correlation functions.

Our main result is described in section~\ref{s:3}, and proved in section~\ref{s:4}.
Section~\ref{s:5} contains discussion of related results.
As a crucial background, we shall review in section~\ref{s:eqt} standard results on the thermodynamic limits of the free energy (the Massieu function) and the entropy, and the equivalence of ensembles in terms of these thermodynamic functions.
In the Appendix, we present (hopefully readable) proofs of these results for completeness.

\section{Equivalence of ensembles for thermodynamic functions}
\label{s:eqt}
Let us start by defining the system we study, and then reviewing standard results on the  equivalence of ensembles for thermodynamic functions in the infinite volume limit \cite{Ruelle,Lima71}.
We shall prove these results in the Appendix.

\paragraph{Model}
We consider a quantum spin system on the $d$-dimensional hypercubic lattice
\eq
\Lambda=\bigl\{(x^{(1)},x^{(2)},\ldots,x^{(d)})\,\bigl|\,x^{(j)}\in\{1,2,\ldots,L\}\bigr\}\subset\bbZ^d
\en
with side length $L$ and with periodic boundary conditions.
The choice of boundary conditions is essential for our equivalence theorem.  
See section~\ref{s:ext}.
We denote by $N=L^d$ the number of sites in the lattice.

For any $x,y\in\Lambda$, we denote by $|x-y|$ the euclidean distance between $x$ and $y$ which takes into account periodic boundary conditions.
For any subsets $X,Y\subset\Lambda$, we define their distance as\footnote{%
We note in passing that this ``distance'' does not satisfy the triangle inequality $\dist(X,Y)\le\dist(X,Z)+\dist(Z,Y)$.
} $\dist(X,Y):=\min_{x\in X, y\in Y}|x-y|$.

We consider a quantum spin system with spin $S\in\{1/2,1,3/2,\ldots\}$ on $\Lambda$.
To be precise we associate  with each site $x\in\Lambda$ a finite dimensional Hilbert space $\calH_x\cong\mathbb{C}^{2S+1}$, and construct the total Hilbert space as $\calH_N=\bigotimes_{x\in\Lambda}\calH_x\cong\mathbb{C}^{(2S+1)^N}$.
For any operator $\hA$ on $\calH_N$, let $\norms{\hA}$ be the operator norm, and $\supp\hA\subset\Lambda$ be its support, i.e., the set of sites on which $\hA$ acts nontrivially.

Let
\eq
\hHN:=\sum_{x\in\Lambda}\hat{h}_x
\lb{Ham}
\en
be the Hamiltonian, which we assume to be translation invariant.
The local Hamiltonian $\hat{h}_x$ is independent of the size $N$, and has a finite support.
We denote by $\kj$ the normalized $j$-th energy eigenstate, i.e., $\hHN\kj=\EN_j\kj$ with $j=1,\dots,\GN:=(2S+1)^N$.
It is easily proved that the minimum and the maximum energy densities
\eq
\umin:=\limN\frac{1}{N}\min\{\EN_1,\ldots,\EN_{\GN}\},\quad
\umax:=\limN\frac{1}{N}\max\{\EN_1,\ldots,\EN_{\GN}\}
\lb{uminmax}
\en
are well-defined.
See Appendix~\ref{a:e}.

\paragraph{Thermodynamic functions}
Let $\beta\in(0,\infty)$ be the inverse temperature.
We define the Massieu function\footnote{
\label{fn:Massieu}
Of course one has $\varphi_N(\beta)=-\beta f_N(\beta)$, where $f_N(\beta)$ is the more standard Helmholtz free energy.
But it is much more convenient to use $\varphi$ than $f$ when one compares the canonical and the microcanonical ensembles.
} by
\eq
\varphi_N(\beta):=\frac{1}{N}\log Z_N(\beta),
\lb{massieu}
\en
where 
\eq
Z_N(\beta):=\operatorname{Tr}[e^{-\beta\hHN}]=\sum_{j=1}^{\GN}e^{-\beta \EN_j}
\lb{ZN}
\en
is the partition function.
It can be easily proved that the limit
\eq
\varphi(\beta):=\limN\varphi_N(\beta)
\lb{philim}
\en
exists, and is convex in $\beta$.
The convergence is uniform in any closed interval in $(0,\infty)$.
See Appendix~\ref{a:f}.

Let the number of states $\ON(U)$ be the number of $j$ such that $\EN_j\le U$.
The entropy density at energy density $u\in(\umin,\umax)$ is defined as
\eq
\sigma_N(u):=\frac{1}{N}\log\ON(Nu).
\lb{sigma1}
\en
It is also provable that the limit
\eq
\sigma(u)=\limN\sigma_N(u)
\lb{sigma2}
\en
exists, and is nondecreasing and concave in $u$.
The convergence is uniform in any closed interval in $(\umin,\umax)$.
See Appendix~\ref{a:s}.

In a quantum spin system, there exists an energy density $\umid$, and the entropy density $\sigma(u)$ is strictly increasing in the interval $(\umin,\umid)$, and stays constant in the interval $[\umid,\umax)$.
The latter interval corresponds to the region with ``negative temperature''.

\paragraph{Equivalence of thermodynamic functions}
The equivalence of ensembles states that the Massieu function $\varphi(\beta)$ and the entropy density $\sigma(u)$ are related by the Legendre transformation.
More precisely, for any $\beta_0\in(0,\infty)$, it can be proved that
\eq
\varphi(\beta_0)=\max_u\{\sigma(u)-\beta_0u\}.
\lb{Legendre2}
\en
Here the maximum is attained at (not necessarily unique) $u$ in the range $(\umin,\umid)$.
For $u_0\in(\umin,\umid)$, we can prove the inverse relation
\eq
\sigma(u_0)=\min_\beta\{\varphi(\beta)+\beta u_0\}.
\lb{Legendre}
\en
See Appendix~\ref{s:sigmaD} for the proofs.

Suppose that the maximum in the right-hand side of \rlb{Legendre2} is attained at a unique value of $u$, which we write $u_0$.
This is indeed the case, unless there is a phase coexistence at $\beta_0$.
Then the Legendre transformation \rlb{Legendre2} reads
\eq
\varphi(\beta_0)=\sigma(u_0)-\beta_0 u_0,
\lb{Legendre3}
\en
which is nothing but the familiar relation $F=U-TS$ written in terms of the Massieu function.
See footnote~\ref{fn:Massieu}.

\paragraph{Different expression for the entropy}
It is also common in physics literature to define the entropy not by the number of states $\ON(U)$, but by the number of states in a certain energy interval.
Let us examine such a formulation.

For any energy $U$ and energy width $\Delta_N>0$, let
\eq
D_N(U,\Delta_N):=\ON(U)-\ON(U-\Delta_N),
\lb{DN}
\en
which is the number of energy eigenstates $j$ such that $U-\Delta_N<E_j\le U$.

Note that \rlb{sigma1} and \rlb{sigma2} roughly implies $\ON(U)\sim\exp[N\sigma(U/N)]$ for large $N$.
For $U$ such that $\sigma(u)$ is strictly increasing at $u=U/N$, this suggests that $D_N(U,\Delta_N)\sim\exp[N\sigma(U/N)]$ for large $N$ and not too small $\Delta_N$.
We thus expect that, for $u\in(\umin,\umid)$, the entropy density $\sigma(u)$ defined by  \rlb{sigma1} and \rlb{sigma2} is also written as
\eq
\sigma(u)=\limN\frac{1}{N}\log D_N(Nu,\Delta_N),
\lb{sigma3}
\en
if we choose of the width $\Delta_N$ properly.

We first see that, in general, the width $\Delta_N$ cannot be a constant (independent of $N$) or anything smaller.
This is because there are models whose energy eigenvalues are integer multiples of a fixed constant\footnote{%
The (classical) Ising model with uniform interactions is an example.
}, and, in such cases, \rlb{sigma3} can never be valid with a constant $\Delta_N$.
A standard and natural choice of the width is $\Delta_N=N\Di u$, where $\Di u$ is a (usually very small) fixed width of the energy density.
With this choice, \rlb{sigma3} is verified easily as follows.

\bigskip\noindent
{\em Proposition}\/: Suppose that $\sigma(u)-\sigma(u-\Di u)>0$ for some $u$ and $\Di u>0$.
Then \rlb{sigma3} holds for any $\Delta_N$ such that $\Delta_N\ge N\Di u$.

\bigskip\noindent
{\em Proof}\/: We fix  $u$ and $\Di u>0$, and let $\varepsilon=\{\sigma(u)-\sigma(u-\Di u)\}/3$.
From the convergence \rlb{sigma2}, there exists $N_0$ such that $|\sigma(u)-\sigma_N(u)|\le\varepsilon$ and $|\sigma(u-\Di u)-\sigma_N(u-\Di u)|\le\varepsilon$ for any $N\ge N_0$.
This implies $\sigma_N(u)-\sigma_N(u-\Di u)\ge\varepsilon$ for any $N\ge N_0$.
From the definition \rlb{sigma1}, we find
\eq
\frac{\Omega_N\bigl(N(u-\Di u)\bigr)}{\Omega_N(Nu)}=\exp\bigl[
N\{\sigma_N(u-\Di u)-\sigma_N(u)\}
\bigr]\le e^{-N\varepsilon},
\en
for any $N\ge N_0$.
Therefore for any $\Delta_N\ge N\Di u$, one has
\eq
\Omega_N(Nu)\ge D_N(Nu,\Delta_N)\ge D_N(Nu,N\Di u)\ge \Omega_N(Nu)\,(1-e^{-N\varepsilon}),
\en
again for any $N\ge N_0$, which implies \rlb{sigma3}.~\qedm

\bigskip
It is likely that  \rlb{sigma3} is valid whenever the width $\Delta_N$ diverges as $N\up\infty$.
But we do not see how to  to prove it.
In Appendix~\ref{s:sigmaD}, we prove a close (but different) relation \rlb{DUND} with a ``fine tuned'' energy range, which is valid for much smaller energy width.

\section{Main result and assumptions}
\label{s:3}
We shall state our main result on the local equivalence of the canonical and the microcanonical ensembles.

\paragraph{Some more definitions}
For any operator $\hA$ on $\calH_N$, we define its canonical average as
\eq
\sbkt{\hA}^{\rm can}_{N,\beta}
:=\frac{\operatorname{Tr}[\hA\,e^{-\beta\hHN}]}{Z_N(\beta)}
=\frac{1}{Z_N(\beta)}\sum_{j=1}^{\GN}\jAj e^{-\beta \EN_j},
\lb{can}
\en
where $Z_N(\beta)$ is the partition function \rlb{ZN}, 
and its microcanonical average (with energy width $\Delta_N$) as\footnote{%
In \rlb{mc}, $\EN_j\in(U-\Delta_N,U]$ specifies the condition for the sum.
We use the same notation throughout the paper.
}
\eq
 \sbkt{\hA}^{\rm mc}_{N,U,\Delta_N}:=\frac{1}{D_N(U,\Delta_N)}
 \sumtwoone{j=1}{\GN}{\bigl(\EN_j\in(U-\Delta_N,U]\bigr)}\jAj,
 \lb{mc}
\en
where the normalization factor $D_N(U,\Delta_N)$ is defined in \rlb{DN}.
For the energy width $\Delta_N$, we only require that $\Delta_N\ge\delta$, where $\delta>0$ is an arbitrary constant that we shall fix throughout the present paper.
The standard choice $\Delta_N=N\Di u$ is of course covered, and the generalized microcanonical average (with extremely large $\Delta_N$) defined by\footnote{
Although this version of microcanonical average is not standard in physics literature, it works equally well as any of the standard versions.
It is also compatible with the definition \rlb{sigma1} of the entropy density.
}
\eq
 \sbkt{\hA}^{\rm gmc}_{N,U}:=\frac{1}{\ON(U)}
 \sumtwoone{j=1}{\GN}{(\EN_j\le U)}\jAj,
 \lb{gmc}
\en
is also allowed.

Note that we here allow the width $\Delta_N$ to be a constant independent of $N$.
This may seem to be contrary to what we have remarked above, but  our equivalence theorem is valid for such small energy width $\Delta_N$ because we ``fine tune'' the energy range of the microcanonical ensemble by defining $U_N(\bz)$ below.
For a model whose energy eigenvalues are integer multiples of a constant (such as the Ising model), our prescription automatically choses an energy range which includes a highly degenerate energy eigenvalue.\footnote{%
It is expected that such a fine tuning is not necessary for generic models in which energy eigenvalues are distributed almost continuously.
But we still do not have any rigorous results in that direction.
}

Let us fix an inverse temperature $\beta_0\in(0,\infty)$.
For each $N$ we define the energy $U_N(\bz)$ corresponding to $\bz$ as the energy $U$ that maximizes $D_N(U,\delta)\,e^{-\beta U}$.
We do not need to assume that the maximizer is unique.

Under the assumption that the right-hand side of \rlb{Legendre2} takes its maximum at a single value of $u$, which we write $u_0$, we can prove that 
\eq
\limN\frac{U_N(\beta_0)}{N}=u_0.
\en
See Proposition in Appendix~\ref{s:sigmaD}.
Interestingly we do not make use of this fact in the proof of the main result.

\paragraph{Main results}
For $\ell\le L/4$, we take a  $d$-dimensional  $\ell\times\cdots\ell$ hypercubic lattice $B$ within $\Lambda$.
We denote the number of sites in the block $B$ as $n=\ell^d$.

We can now state the main result of the present paper.

\bigskip\noindent
{\em Theorem}\/: Suppose that Assumptions I and II, which we shall describe below, are valid.
This is guaranteed when $\bz$ is sufficiently small.
There exist a positive constant\footnote{
In the present paper, a constant may depend on the dimension $d$, the Hamiltonian, the fixed inverse temperature $\bz$, and the fixed minimum width $\delta$, but not on  the operators, or the sizes $N$, $n$.
} $C$ and a positive function\footnote{
$N_0(\ep)$ may depend on  $d$, the Hamiltonian, $\bz$, and $\delta$, but not on  the operators, or the sizes $N$, $n$.
} $N_0(\ep)$ of $\ep\in(0,1/2)$ which diverges as $\ep\downarrow0$.
Take an arbitrary self-adjoint operator $\hA$ whose support is contained in the block $B$.
Then for any $\ep\in(0,1/2)$ and $n$, we have
\eq
\abs{\bktcan{\hA}-\sbkt{\hA}^{\rm mc}_{N,U_N(\bz),\Delta_N}}\le C\rbk{\frac{n}{N^{(1/2)-\ep}}}^{1/2}\,\norms{\hA},
\lb{main}
\en
for arbitrary $N$ such that $N\ge N_0(\ep)$.

\bigskip
The theorem states that, whenever $n/N^{(1/2)-\ep}$ is sufficiently small, the canonical and the microcanonical expectation values almost coincide.
This is essentially the theorem of Brandao and Cramer \cite{BrandaoCramer} for the equivalence of ensembles in terms of local states within a finite volume.\footnote{
We note that our version is weaker in some aspects.
For example we can only treat the microcanonical ensemble with the energy $U_N(\bz)$ while Brandao and Cramer allow small changes in the energy.
This is a technical limitation of our elementary proof.
On the other hand we are able to treat larger block $B$ and smaller energy width $\Delta_N$ compared with  \cite{BrandaoCramer}.
}

It is well-known that, for $\bz$ satisfying Assumption~I, one has
\eq
\limN\bktcan{\hA}=\rho^{\rm KMS}_{\bz}[\hA],
\en
for any local self-adjoint operator $\hA$ (i.e., a self-adjoint operator on $\calH_N$ for sufficiently large $N$), where $\rho^{\rm KMS}[\cdot]$ is the unique infinite volume equilibrium state (called the KMS state) at $\bz$ \cite{BratteliRobinson1,BratteliRobinson2}.
Then, by fixing $n=\ell^d$ and letting $N=L^d$ tend to infinity in \rlb{main}, we get the following infinite volume version \cite{MartinLof,Georgii,Lima72,MuellerAdlamMasanesWiebe} of the equivalence theorem.

\bigskip\noindent
{\em Corollary}\/:
Under the same conditions as the above theorem, we have
\eq
\limN\sbkt{\hA}^{\rm mc}_{N,U_N(\bz),\Delta_N}=\rho^{\rm KMS}_{\bz}[\hA],
\en
for any local self-adjoint operator $\hA$.

\paragraph{Assumptions}
Let us state the two assumptions required for the theorem.
We note that both the assumptions have been proved for an arbitrary model in any dimension $d$, provided that $\bz$ is sufficiently small\footnote{
Unfortunately it is not easy to locate references where the exact statements appear.
But all the properties have been (at least implicitly) proved, e.g., in \cite{Park,Simon,FroelichUeltschi}.
}.
The proofs are based on the cluster expansion method.
See, e.g., \cite{Park,Simon,FroelichUeltschi}.
For $d=1$, we expect that both the assumptions can be verified for any $\bz\in(0,\infty)$ by using ``classical'' works on quantum spin chains such as \cite{Araki}, but we have not checked the details.\footnote{
Eq. (8.56) of \cite{Araki} establishes the desired exponential decay with better constant than in \rlb{expdecay}, but for the infinite system.
}

\bigskip\noindent
{\em Assumption I}\/:
There are positive constants  $C_1$ and $\xi$.
For any $N$ and arbitrary self-adjoint operators $\hA$ and $\hB$ on $\calH_N$, one has
\eq
\abs{\bktcan{\hA\hB}-\bktcan{\hA}\bktcan{\hB}}\le
C(\hA,\hB)\,\exp\biggl[-\frac{\dist(\supp\hA,\supp\hB)}{\xi}\biggr]
\lb{expdecay}
\en
with\footnote{
By $|S|$ we denote the number of elements in a set $S$.
}
\eq
C(\hA,\hB)=
C_1\,\norms{\hA}\,\norms{\hB}\,\bigl|\operatorname{supp}\hA\bigr|\,\bigl|\operatorname{supp}\hB\bigr|.
\en

\bigskip\noindent
{\em Assumption II}\/:
There are $\beta_1$ and $\beta_2$ such that $\beta_1<\bz<\beta_2$, and the following two properties are valid.
There is a positive constant $C_2$, and one has
\eq
|\varphi_N(\beta)-\varphi(\beta)|\le \frac{C_2}{N},
\lb{phiconv}
\en
for any $\beta\in[\beta_1,\beta_2]$ and $N$.
The Massieu function $\varphi(\beta)$ is twice continuously differentiable, and satisfies $\varphi''(\beta)\ge c_0$ with a constant $c_0>0$ in the interval $[\beta_1,\beta_2]$.

\bigskip

Assumption~I states that any two-point truncated correlation function for the canonical ensemble exhibits exponential decay.
This property is expected to be valid for any temperatures higher than the critical temperature, but a general proof is still lacking.
When $\bz$ is sufficiently small, it can be proved rigorously by using the cluster expansion technique.
See, e.g., Theorem~3.2 of \cite{FroelichUeltschi}.

The first statement of Assumption~II concerns the speed of convergence to the infinite volume limit of the Massieu function.
Indeed it can be proved easily by using standard methods in rigorous statistical mechanics \cite{Ruelle,Simon} that 
\eq
|\varphi_N(\beta)-\varphi(\beta)|\le \frac{K(\beta)}{L},
\lb{phiconv2}
\en
for any $\beta$, with $K(\beta)<\infty$.
See the discussion after \rlb{TrTrTre} in Appendix~\ref{s:TDL}.
Since $N=L^d$, this guarantees \rlb{phiconv} only for $d=1$.
If we restrict ourselves to sufficiently small $\beta$, then the cluster expansion technique allows us to establish the very rapid convergence\footnote{
Again we are not able to locate a single suitable reference.
But one easily finds this property by examining the cluster expansion of the free energy for a translation-invariant system, either classical or quantum, with periodic boundary conditions.
}
\eq
|\varphi_N(\beta)-\varphi(\beta)|\le K'(\beta) e^{-\gamma(\beta)\,L},
\en
for any $d$, with $\gamma(\beta)>0$ and $K'(\beta)<\infty$.
This is certainly more than enough to justify \rlb{phiconv}.
We note that this rapid convergence is provable (and valid) only for models with periodic boundary conditions.
With other boundary conditions, the difference $|\varphi_N(\beta)-\varphi(\beta)|$ may be of the order of $1/L$.

The second statement of Assumption~II basically states that the model has a positive specific heat.
The statement has been proved by using the cluster expansion technique when $\beta$ is sufficiently small.
Indeed it has been shown that $\varphi(\beta)$ is analytic for small $|\beta|$.

\section{Proof}
\label{s:4}

\subsection{Lemmas and the proof of Theorem}
We state two essential lemmas, and prove the theorem.

Let us make $M$ copies $B_1,\ldots,B_M$ of the block $B$, and embed them into $\Lambda$ in such a manner that $\dist(B_i,B_j)\ge\ell$ holds for any $i\ne j$.
One can take $M$ so as to satisfy
\eq
M\ge\left[\frac{L}{2\ell}\right]^d\ge\frac{1}{4^d}\frac{N}{n},
\lb{M}
\en
where $[x]$ is the largest integer that does not exceed $x$, and we used $\ell\le L/4$ to get the final lower bound.

Take an arbitrary self-adjoint operator $\hA$ with $\supp\hA\subset B$, and let $\hA_i$ be an exact translational copy of $\hA$ whose support is contained in $B_i$.

\bigskip\noindent
{\em Lemma I}\/:
Under Assumption I, there is a constant $C_3$, and we have
\eq
\bkt{\biggl\{\frac{1}{M}\sum_{i=1}^M\Bigl(
\hA_i-\bktcan{\hA_i}
\Bigr)\biggr\}^2}^{\rm can}_{N,\beta_0}\le C_3\frac{\norms{\hA}^2}{M},
\lb{L1}
\en
for any $N$, $n$, and $M$.

\bigskip\noindent
{\em Lemma II}\/:
Under Assumption II, there exists a positive constant $C_4$.
For an arbitrary non-negative operator $\hC$ on $\Lambda$, we have
\eq
\sbkt{\hC}^{\rm mc}_{N,U_N(\beta_0),\Delta_N}\le C_4\,N^{(1/2)+\ep}\,\bktcan{\hC},
\lb{L2}
\en
for arbitrary $N$ such that $N\ge N_0(\ep)$, provided that $\Delta_N\ge\delta$.

\bigskip
Let us prove Theorem by assuming Lemmas I and II.
By combining \rlb{L2} and \rlb{L1}, one finds
\eqa
\Biggl\langle\biggl\{\frac{1}{M}\sum_{i=1}^M&\Bigl(
\hA_i-\bktcan{\hA_i}
\Bigr)\biggr\}^2\Biggr\rangle^{\rm mc}_{N,U_N(\beta_0),\Delta_N}
\nl
&\le
C_4\,N^{(1/2)+\ep}\bkt{\biggl\{\frac{1}{M}\sum_{i=1}^M\Bigl(
\hA_i-\bktcan{\hA_i}
\Bigr)\biggr\}^2}^{\rm can}_{N,\beta_0}
\nl&\le 
C_3\,C_4\frac{N^{(1/2)+\ep}}{M}\norms{\hA}^2
\le4^d\,C_3\,C_4\frac{n}{N^{(1/2)-\ep}}\norms{\hA}^2,
\lb{main0}
\ena
where we used \rlb{M}.
Next note that, for any self-adjoint operator $\hX$ and $a\in\bbR$, one has 
\eq
\bbkt{\{\hX-(\sbkt{\hX}+a)\}^2}=\bbkt{\{\hX-\sbkt{\hX}\}^2}+a^2\ge a^2,
\lb{Xa}
\en
where $\sbkt{\cdots}$ denotes an arbitrary average.
By using this (trivial) inequality and the translation invariance, we find that the left-hand side of \rlb{main0} is bounded from below by $(\bktcan{\hA}-\sbkt{\hA}^{\rm mc}_{N,U_N(\beta_0),\Delta_N})^2$.
Thus the desired bound \rlb{main} has been proved.

\bigskip\noindent
{\em Remark}\/:
Let us note a simple implication of Lemma~II.
Let $\hat{a}_x$ be the translation of a self-adjoint operator $\hat{a}_o$ with a finite support, and consider the corresponding macroscopic operator $\hA_N:=N^{-1}\sum_{x\in\Lambda}\hat{a}_x$.
When Assumption~I is valid, we obtain the standard estimate
\eq
\bkt{\bigl(\hA_N-\bktcan{\hA_N}\bigr)^2}^{\rm can}_{N,\bz}\le\frac{(\text{const.})}{N},
\en
directly from the assumed bound \rlb{expdecay}.
This means that the fluctuation of $\hA_N$ in the canonical ensemble is of $O(N^{-1/2})$.
Noting that $\bbkt{\{\hX-\sbkt{\hX}\}^2}\le\bbkt{\{\hX-(\sbkt{\hX}+a)\}^2}$ for any $a\in\bbR$, where we used the same notation as in \rlb{Xa}, we find
\eqa
\Bigl\langle\bigl(\hA_N-&\sbkt{\hA_N}^{\rm mc}_{N,U_N(\beta_0),\Delta_N}\bigr)^2\Bigr\rangle^{\rm mc}_{N,U_N(\beta_0),\Delta_N}
\le
\bkt{\bigl(\hA_N-\bktcan{\hA_N}\bigr)^2}^{\rm mc}_{N,U_N(\beta_0),\Delta_N}
\nl
&\le
 C_4\,N^{(1/2)+\ep}\bkt{\bigl(\hA_N-\bktcan{\hA_N}\bigr)^2}^{\rm can}_{N,\bz}\le\frac{(\text{const.})}{N^{(1/2)-\ep}},
\ena
where we used the inequality \rlb{L2}.
This implies that
\eq
\limN\Bigl\langle\bigl(\hA_N-\sbkt{\hA_N}^{\rm mc}_{N,U_N(\beta_0),\Delta_N}\bigr)^2\Bigr\rangle^{\rm mc}_{N,U_N(\beta_0),\Delta_N}=0,
\en
i.e., the fluctuation of $\hA_N$ in the microcanonical ensemble vanishes in the thermodynamic limit.

\subsection{Proof of Lemma I}
The proof is straightforward.
Note first that
\eq
\bkt{\biggl\{\frac{1}{M}\sum_{i=1}^M\Bigl(
\hA_i-\bktcan{\hA_i}
\Bigr)\biggr\}^2}^{\rm can}_{N,\beta_0}
\le\frac{1}{M^2}\sum_{i=1}^M\sum_{j=1}^M
\abs{\bktcan{\hA_i\hA_j}-\bktcan{\hA_i}\bktcan{\hA_j}}.
\en
We shall show for each $i$ that
\eq
\sum_{j=1}^M
\abs{\bktcan{\hA_i\hA_j}-\bktcan{\hA_i}\bktcan{\hA_j}}
\le  C_3\norms{\hA}^2,
\lb{L11}
\en
which implies \rlb{L1}.

As for the term in the left-hand side of \rlb{L11} with $i=j$, we simply use the bound $\abs{\bktcan{\hA_i\hA_i}-\bktcan{\hA_i}\bktcan{\hA_i}}\le2\norms{\hA}^2$.
For terms with $i\ne j$, we use the assumption \rlb{expdecay} to see 
\eq
\abs{\bktcan{\hA_i\hA_j}-\bktcan{\hA_i}\bktcan{\hA_j}}
\le C_1 n^2 \norms{\hA}^2 e^{-\dist(B_i,B_j)/\xi}.
\en
The sum over $j$ of the right-hand side, which gives an upper bound of the left-hand side of \rlb{L11}, converges because of the exponential decay.
Although the resulting sum depends on the choice of $n=\ell^d$, it converges to 0 as $n\up\infty$.
We have thus proved the desired bound \rlb{L11} with a constant $C_3$ independent of $n$.

\subsection{Proof of Lemma II}
The proof of Lemma II is the core of the present paper.
Let us define the energy density $u_0$ which corresponds to the fixed inverse temperature $\bz$ by
\eq
u_0:=-\varphi'(\bz),
\lb{u0}
\en
where we noted that $\varphi(\beta)$ is differentiable at $\beta=\bz$ because of Assumption~II.
Then  one finds, from \rlb{Legendre}, that
\eq
\sigma(u_0)=\varphi(\bz)+\bz u_0,
\lb{su0}
\en
and, from \rlb{Legendre2}, that $\bz=\sigma'(u_0)$.

We first observe, by using the definitions \rlb{mc} and \rlb{can}, that
\eqa
\sbkt{\hC}^{\rm mc}_{N,U_N(\beta_0),\Delta_N}&=\frac{1}{D_N(U_N(\beta_0),\Delta_N)}
\sumtwo{j}{\bigl(\EN_j\in(U_N(\beta_0)-\Delta_N,U_N(\beta_0)]\bigr)}\bra{j}\hC\kj
\nl
&\le\frac{e^{\beta_0\,U_N(\beta_0)}}{D_N(U_N(\beta_0),\Delta_N)}\sumtwo{j}{\bigl(\EN_j\in(U_N(\beta_0)-\Delta_N,U_N(\beta_0)]\bigr)}\bra{j}\hC\kj\,e^{-\beta_0\EN_j}
\nl
&\le\frac{Z_N(\beta_0)}{D_N(U_N(\beta_0),\Delta_N)\,e^{-\beta_0\,U_N(\beta_0)}}\,\bktcan{\hC}
\nl
&\le\frac{Z_N(\beta_0)}{D_N(U_N(\beta_0),\delta)\,e^{-\beta_0\,U_N(\beta_0)}}\,\bktcan{\hC}\,,
\ena
where we used $\Delta_N\ge\delta$ to get the final inequality.
We shall show below that 
\eq
\frac{Z_N(\beta_0)}{D_N(U_N(\beta_0),\delta)\,e^{-\beta_0\,U_N(\beta_0)}}\le C_4\,N^{(1/2)+\ep},
\lb{ZOe}
\en
for any $N$ such that $N\ge N_0(\ep)$.
This implies the desired \rlb{L2}.
We note that the bound \rlb{ZOe} with the right-hand side replaced by $(\text{const.})\,N$ is well-known and easily proved as in \rlb{ZN<} in Appendix~\ref{s:sigmaD}.
It was essential for the present poof to reduce the power of $N$.
We believe the power can never be less than 1/2.\footnote{
To see this we examine the standard heuristic estimate of the partition function.
Let us approximate the sum by an integral as $Z_N(\bz)\simeq\int dE\,\tilde{D}_N(E)\,e^{-\bz E}$ with $\tilde{D}_N(E)=d\ON(E)/dE$.
Since $\tilde{D}_N(E)$ behaves essentially like $\ON(E)$, we have $\tilde{D}_N(E)/\tilde{D}_N(u_0N)\simeq\exp[N\{\sigma(u)-\sigma(u_0)\}]$, where  $u=E/N$.
Because $\sigma'(u_0)=\bz$, expansion to the second order in $(u-u_0)$ yields \newline$\sigma(u)-\sigma(u_0)-\beta u\simeq-\bz u_0-\alpha(u-u_0)^2$. 
Then we find \newline$Z_N(\bz)\simeq \tilde{D}_N(u_0N)\,e^{-\bz u_0N}\,N\int du\,e^{-N\alpha(u-u_0)^2}=\sqrt{2\pi N/\alpha}\,\tilde{D}_N(u_0N)\,e^{-\bz u_0N}$, which should be compared with \rlb{ZOe}.
}

For $u_0$ defined in \rlb{u0}, one can easily show (see below) for any $\Du>0$ the standard large-deviation type upper bound
\eq
\Biggl\langle\hat{P}\biggl[\Bigl|\frac{\hHN}{N}-u_0\Bigr|\ge\Du\biggr]\Biggr\rangle^{\rm can}_{N,\beta_0}\le\exp\bigl[-N\min\{\psi(\Du),\psi(-\Du)\}+2C_2\bigr],
\lb{LD}
\en
where $\hat{P}[\cdots]$ is the orthogonal projection onto the specified subspace, and the large deviation function $\psi(x)$ is given by 
\eq
\psi(x):=\sigma(u_0)-\sigma(u_0+x)+\beta_0x.
\lb{psi}
\en

Because, by Assumption~II, the Massieu function $\varphi(\beta)$ is twice continuously differentiable and strictly convex  in $[\beta_1,\beta_2]$, its Legendre transform $\sigma(u)$ is strictly concave and twice continuously differentiable in a finite open interval containing $u_0$.
Then the large deviation function $\psi(x)$ is of course strictly convex and twice continuously differentiable in the same interval.
Noting that $\psi(0)=0$ and $\psi'(0)=0$ because $\sigma'(u_0)=\beta_0$, we see that there are positive constants $\alpha$, $x_0$, and $\psi(x)\ge\alpha\,x^2$ for any $x\in(-x_0,x_0)$.

Let us now choose
\eq
\Du=\frac{N^{(1/2)+\ep}}{N}=N^{-(1/2)+\ep},
\en
and suppose that $N$ is large enough so that $\Du<x_0$.
Then \rlb{LD} implies
\eq
\biggl\langle\hat{P}\Bigl[\bigl|\hHN-N\,u_0\bigr|\ge N^{(1/2)+\ep}\Bigr]\biggr\rangle^{\rm can}_{N,\beta_0}\le e^{-\alpha N^{2\ep}+2C_2}.
\lb{LD3}
\en
We now choose the function $N_0(\ep)$ so that (already mentioned) $\Du<x_0$ and
\eq
e^{-\alpha N^{2\ep}+2C_2}\le \frac{1}{2}
\lb{N0det}
\en
hold for any $N$ such that $N\ge N_0(\ep)$.
Note that $N_0(\ep)$ grows as $N_0(\ep)\sim(\text{const.})^{1/\ep}$ as $\ep\downarrow0$.

We now define
\eq
\tilde{Z}_N(\beta_0):=\sumtwo{j}{\bigl(|\EN_j-Nu_0|<N^{(1/2)+\ep}\bigr)}
e^{-\beta_0\EN_j},
\en
and observe that, for $N\ge N_0(\ep)$,
\eq
1-\frac{\tilde{Z}_N(\beta_0)}{Z_N(\beta_0)}=\biggl\langle\hat{P}\Bigl[\bigl|\hHN-N\,u_0\bigr|\ge N^{(1/2)+\ep}\Bigr]\biggr\rangle^{\rm can}_{N,\beta_0}\le e^{-\alpha N^{2\ep}+2C_2}\le\frac{1}{2},
\en
which implies
\eq
Z(\beta_0)\le2\,\tilde{Z}_N(\beta_0).
\lb{ZZ}
\en

For each $N$, we take a sequence $U_0,U_1,\ldots,U_M$ of energies such that (i)~$U_\nu=U_N(\bz)$ for one $\nu$, (ii)~$U_{\nu}-U_{\nu-1}=\delta$ for $\nu=1,\ldots,M$, and (iii)~$(U_0,U_M)\ni E_j$ for all $j=1,\ldots,\GN$.
Then note by definition that
\eqa
\tilde{Z}_N(\beta_0)&\le\sumtwo{\nu\in\bbZ}{\bigl(|U_\nu-Nu_0|\le \delta+N^{(1/2)+\ep}\bigr)}
D_N(U_\nu,\delta)\,e^{-\bz U_{\nu-1}}
\nl
&\le e^{\beta_0\delta}\rbk{\frac{2N^{(1/2)+\ep}}{\delta}+2}\,\max_\nu D_N(U_\nu,\delta)\,e^{-\beta_0U_\nu}
\nl
&= e^{\beta_0\delta}\rbk{\frac{2N^{(1/2)+\ep}}{\delta}+2}\,D_N(U_N(\beta_0),\delta)\,e^{-\beta_0U_N(\beta_0)},
\lb{tdbound}
\ena
where we recalled that $U_N(\beta_0)$ is the energy $U$ that maximizes $D_N(U,\delta)\,e^{-\bz U}$.
This, with \rlb{ZZ}, implies the desired bound \rlb{ZOe}.

Finally we prove the large-deviation type upper bound \rlb{LD} for completeness.
Note that for any $\lambda>0$ one has
\eqa
\Biggl\langle\hat{P}\biggl[\frac{\hHN}{N}-u_0\ge\Du\biggr]\Biggr\rangle^{\rm can}_{N,\beta_0}&\le
\langle e^{\lambda\{\hHN-N(u_0+\Du)\}}\rangle^{\rm can}_{N,\beta_0}
\nl
&=e^{N\{\varphi_N(\beta_0-\lambda)-\varphi_N(\beta_0)-\lambda(u_0+\Du)\}}
\nl
&\le e^{N\{\varphi(\beta_0-\lambda)-\varphi(\beta_0)-\lambda(u_0+\Du)\}+2C_2},
\lb{LD2}
\ena
where we used \rlb{phiconv} to get the final expression.
We now observe, by writing $\lambda=\bz-\tilde{\beta}$, that
\eqa
\inf_{\lambda>0}\{\varphi(\beta_0-\lambda)-&\varphi(\beta_0)-\lambda(u_0+\Du)\}
\nl&
=\inftwo{\tilde{\beta}}{(\tilde{\beta}<\bz)}\{\varphi(\tilde{\beta})-\varphi(\beta_0)-(\bz-\tilde{\beta})(u_0+\Du)\}
\nl
&=\inftwo{\tilde{\beta}}{(\tilde{\beta}<\bz)}\{\varphi(\tilde{\beta})+\tilde{\beta}(u_0+\Du)\}-\{\varphi(\beta_0)+\bz u_0\}-\bz\Du
\nl
&=\sigma(u_0+\Du)-\sigma(u_0)-\bz\Du,
\ena
where we used \rlb{Legendre} and \rlb{su0}.
By substituting this back into \rlb{LD2}, we get
\eq
\Biggl\langle\hat{P}\biggl[\frac{\hHN}{N}-u_0\ge\Du\biggr]\Biggr\rangle^{\rm can}_{N,\beta_0}
\le e^{-N\psi(\Du)+2C_2},
\en
with $\psi(x)$ defined in \rlb{psi}.

Bounding $\langle\hat{P}[({\hHN}/{N})-u_0\le-\Du]\rangle^{\rm can}_{N,\beta_0}$ in a similar manner, we get \rlb{LD}.

\section{Related issues}
\label{s:5}
\subsection{Some extensions}
\label{s:ext}

\paragraph{Non-local operators}
In the main theorem, we considered an operator $\hA$ whose support is contained in a single  block $B$ which is much smaller than the whole lattice.
It is clear from the proof, however, that the locality is not essential.
What is really important is the possibility to make $M$ translational copies of $\hA$ in such a manner that the supports are sufficiently far away from each other.
We can therefore treat non-local $\hA$ by the same method provided that its support is small enough so that we can distribute many copies over the lattice.

As a typical example, take self-adjoint operators $\hat{a}_x$ and $\hat{b}_y$, which acts strictly on the local Hilbert space $\calH_x$ and $\calH_y$, respectively, and set $\hA=\hat{a}_x\hat{b}_y$.
Although this operator is not covered by the main theorem (unless $x$ and $y$ are contained in a single block), one can repeat the proof by considering the translation $\hat{a}_{x+z_i}\hat{b}_{y+z_i}$ with $i=1,\ldots,M$, requiring that support do not overlap (i.e., $2M$ sites $x+z_1,y+z_1,x+z_2,\ldots,y+z_M$ are all distinct).
In this case one can choose $M\propto N$, and get the following extension of the main theorem.

\bigskip\noindent
{\em Proposition}\/: Under the same assumption as the main theorem, there exists a constant $\tilde{C}$.
For any strictly local self-adjoint operators $\hat{a}_x$, $\hat{b}_y$, and  $\ep\in(0,1/2)$, we have
\eq
\abs{\bktcan{\hat{a}_x\hat{b}_y}-\sbkt{\hat{a}_x\hat{b}_y}^{\rm mc}_{N,U_N(\bz),\Delta_N}}\le \tilde{C}\,N^{-\frac{1}{4}+\frac{\ep}{2}}\,\norms{\hat{a}_x}\,\norms{\hat{b}_y},
\lb{NC}
\en
for arbitrary $N$ such that $N\ge N_0(\ep)$.
There are no restrictions on the location of the sits $x,y\in\Lambda$.

\bigskip
Recall that we can prove the theorem (and the above proposition) only for sufficiently small $\bz$, where the truncated canonical correlation function $\bktcan{\hat{a}_x\hat{b}_y}-\bktcan{\hat{a}_x}\bktcan{\hat{b}_y}$ decays exponentially in the distance $|x-y|$.
Note, however, that the bound \rlb{NC} does not imply the exponential decay of the corresponding truncated microcanonical correlation function.
This is because, on the right-hand side, there is a small but nonzero term which does not depend on $x$ or $y$.
It is likely that this reflects the property of the microcanonical ensemble that some local operators have small uniform correlation.
See section~\ref{s:wM}.

\paragraph{Other boundary conditions}
We can extend our method to the same short-range translation invariant models with boundary conditions other than the periodic boundary conditions, e.g., the free boundary conditions.
It is known that the thermodynamic functions $\varphi(\beta)$ and $\sigma(u)$ in the infinite volume limit are unchanged.
See Appendix~\ref{s:TDL}.

In such a case, however, expectation values (for a finite $N$) is no longer exactly translation invariant because of various boundary effects.
Thus the left-hand side of the main bound \rlb{main} should be replaced by the average over the $M$ blocks (as is done in \cite{BrandaoCramer}).

The assumptions necessary for the proof are again provable using the cluster expansion technique, except for \rlb{phiconv} about the rapid convergence of the Massieu function.
As we mentioned already such a convergence is never expected except for the periodic boundary conditions, and we must proceed with the slower convergence as in  \rlb{phiconv2}.
In this case, $2C_2$ in \rlb{N0det} is replaced by $(\text{const.})\,N^{(d-1)/d}$, and we are forced to take
\eq
\ep>\frac{1}{2}-\frac{1}{2d}\,,
\en
which leads to weaker bounds.

\subsection{The energy width of the microcanonical ensemble}
\label{s:DN}
Let us make some comments on the width $\Delta_N$ of the microcanonical ensemble.

In the canonical ensemble, the fluctuation of the total energy is in general proportional to $\sqrt{N}$.
This well-known fact seems to have caused misunderstanding that one should also choose the energy width $\Delta_N$ of the microcanonical ensemble to be proportional to $\sqrt{N}$.
This is indeed not at all necessary.

In fact it can be shown that the ``effective energy width'', i.e., the fluctuation of the total energy in the microcanonical ensemble, is in general of order one, no matter how large $\Delta_N$ is.
This reflects the existence of the sharp cutoff in the upper limit of the energy in the microcanonical ensemble. 

To see this we consider a generic quantum spin system in which (unlike the Ising model) the density of states can be regarded a an almost smooth function.
We then use the heuristic estimate (which is closely related to \rlb{sigma3})
\eq
\frac{d\Omega_N(E)}{dE}\simeq (\text{const.}) e^{N\sigma(E/N)},
\en
and approximate the microcanonical average (of a function of $\hHN$) as
\eq
\sbkt{\cdots}^{\rm mc}_{U_0,\Delta_N}\simeq
\frac{\int_{U_0-\Delta_N}^{U_0}dE\,(\cdots)\,\{d\Omega_N(E)/dE\}}
{\int_{U_0-\Delta_N}^{U_0}dE\,\{d\Omega_N(E)/dE\}}
\simeq
\frac{\int_{U_0-\Delta_N}^{U_0}dE\,(\cdots)\,e^{N\sigma(E/N)}}
{\int_{U_0-\Delta_N}^{U_0}dE\,e^{N\sigma(E/N)}},
\lb{mcap}
\en
where we set $U_0=u_0N$ with a fixed energy density $u_0$.
Expanding around $U_0$, we find $N\sigma(E/N)\simeq N\sigma(u_0)+\bz(E-U_0)$, where $\bz=\sigma'(u_0)$.
Substituting this into \rlb{mcap}, and assuming $\bz\Delta_N\gg1$, we can further approximate the microcanonical average as
\eq
\sbkt{\cdots}^{\rm mc}_{U_0,\Delta_N}\simeq
\frac{\int_{-\infty}^{U_0}dE\,(\cdots)\,e^{\bz(E-U_0)}}
{\int_{-\infty}^{U_0}dE\,e^{\bz(E-U_0)}}.
\lb{mcap2}
\en
With this simple formula, one readily computes that
\eqg
\sbkt{\hHN}^{\rm mc}_{U_0,\Delta_N}\simeq U_0-(\bz)^{-1}=U_0-\kT,
\\
\bkt{\bigl(\hHN-\sbkt{\hHN}^{\rm mc}_{U_0,\Delta_N}\bigr)^2}^{\rm mc}_{U_0,\Delta_N}\simeq{\bz}^{-2}=(\kT)^2,
\lb{mcEvar}
\eng
where $\kT:=(\bz)^{-1}$.
This means that, as long as $\Delta_N\gg\kT$, the effective energy width of the microcanonical ensemble is always  $\kT$.

But this does not mean that one always have to set $\Delta_N\gg\kT$.
In fact our theorem shows that, at high temperatures, the canonical and the microcanoical ensembles are equivalent only provided that $\Delta_N$ is not less than an arbitrary constant $\delta>0$, if one ``fine-tunes'' the range of energy.
This suggests that the width $\Delta_N$ can be extremely small, even can be a decreasing function of $N$, if it is guaranteed that there are sufficiently many states in the energy range.

\subsection{Weak uniform correlation in the microcanonical ensemble}
\label{s:wM}
In section~\ref{s:ext}, we suggested that a truncated microcanonical correlation function may not exhibit complete exponential decay even when the canonical counterpart does.

To see this in the simplest context, consider the microcanonical ensemble at the energy $U_0$ with an arbitrary width $\Delta_N$.
The fluctuation of the total energy is written as
\eqa
\bbkt{(\hHN-\sbkt{\hHN}^{\rm mc}_{N,U_0,\Delta_N})^2}^{\rm mc}_{N,U_0,\Delta_N}
&=\bkt{\Bigl\{\sum_{x\in\Lambda}(\hat{h}_x-\sbkt{\hat{h}_x}^{\rm mc}_{N,U,\delta})\Bigr\}^2}^{\rm mc}_{N,U_0,\Delta_N}
\nl&=N\sum_{x\in\Lambda}\bkt{(\hat{h}_o-\sbkt{\hat{h}_o}^{\rm mc}_{N,U_0,\Delta_N})(\hat{h}_x-\sbkt{\hat{h}_x}^{\rm mc}_{N,U_0,\Delta_N})}^{\rm mc}_{N,U_0,\Delta_N}.
\ena
Since the left-hand side is always of order one as we discussed above (see \rlb{mcEvar}), we find that
\eq
\sum_{x\in\Lambda}\bkt{(\hat{h}_o-\sbkt{\hat{h}_o}^{\rm mc}_{N,U_0,\Delta_N})(\hat{h}_x-\sbkt{\hat{h}_x}^{\rm mc}_{N,U_0,\Delta_N})}^{\rm mc}_{N,U_0,\Delta_N}\propto N^{-1}.
\en
Now quite generally the term with $x=o$ in the sum $\bbkt{(\hat{h}_o-\sbkt{\hat{h}_o}^{\rm mc}_{N,U_0,\Delta_N})^2}^{\rm mc}_{N,U_0,\Delta_N}$ is strictly positive and of order one.
This means that the remaining terms should sum up to a negative quantity of order one, which suggests  
\eq
\bkt{(\hat{h}_o-\sbkt{\hat{h}_o}^{\rm mc}_{N,U_0,\Delta_N})(\hat{h}_x-\sbkt{\hat{h}_x}^{\rm mc}_{N,U_0,\Delta_N})}^{\rm mc}_{N,U_0,\Delta_N}\simeq-\frac{(\text{positive const.})}{N},
\en
for $x$ such that $|o-x|\gg\xi$.
In other words there is a uniform negative correlation roughly proportional to $N^{-1}$.
One can easily make the above observation a rigorous statement for a non-interacting system (where $\hat{h}_x$ acts only on $\calH_x$).

\section{Discussion}
We have stated and proved a local version of the equivalence theorem between the canonical ensemble with inverse temperature $\bz$ and the microcanonical ensemble with the corresponding ``fine-tuned'' energy $U_N(\bz)$.
Our equivalence theorem is quite general and applies to any quantum spin system with a short-ranged translation invariant Hamiltonian.

In order to verify the two assumptions of the theorem, however, we still need to assume that the temperature is sufficiently high.
It is interesting to see if the theorem can be extended to an arbitrary temperature higher than the critical temperature, or if there can be analogous statements for low enough temperatures (in dimensions two or higher) where the system exhibits long-range order.

The local equivalence of ensembles had been expected for a long time, and was proved by Brandao and Cramer \cite{BrandaoCramer} by using advanced techniques in quantum information theory.
We here presented a much more elementary and shorter proof which yields stronger result for a restricted setting.
One conclusion of the present paper which was probably not noted before is that the width $\Delta_N$ of the microcanonical ensemble can be taken to be a constant if one fine-tunes the energy range.

We believe that our theorem does not only shed light on the basic structure of equilibrium statistical mechanics, but  can also be used to develop theories for large but finite systems in equilibrium.
In fact the present result was recently used in \cite{IKS}, where the second law of thermodynamics and the fluctuation theorem were formulated and proved for a macroscopic quantum system in a pure state.

\appendix
\section{Proof of standard results}
In the Appendix, we  prove standard results on the thermodynamic limits of thermodynamic functions and on the equivalence of ensembles \cite{Ruelle,Lima71,Lima72} discussed in section~\ref{s:eqt}.
We have decided to include this material because, to our surprise, we were not able to find any compact account which contains a complete proof of  these essential results.
We hope that the reader find the following proof written in modern notation useful.

\subsection{Thermodynamic limits}
\label{s:TDL}
We prove the existence of the thermodynamic limits for the energy \rlb{uminmax}, the free energy density \rlb{philim}, and the entropy density \rlb{sigma2}.
Here we prove slightly weaker statements where only the hypercubic lattices with side length $L=2^n$ are considered, with $n$ being positive integers.\footnote{
We of course believe that the limits exist when $N$ varies over all positive integers.
In fact, as for the energy density and the free energy density, the existence of such limits is easily proved by using, e.g, the technique which makes use of the lattice with side length $M=2^nL$.
See p.~16 of \cite{Griffiths}.
As for the entropy density, the proof may be more tricky, but should be doable by following, e.g., \cite{Ruelle}.
}
Therefore the symbol $\limN$ should always be understood to mean $\limn$ with $N=L^d=2^{nd}$.

\subsubsection{Assumptions}
In the present Appendix we only consider the $d$-dimensional hyper cubic lattice whose side length is $L=2^n$ with some $n=1,2,\ldots$, and mainly study quantum spin models with open boundary conditions.
We thus change the notation, and denote by $\hH_n$ and $\hHp_n$ the Hamiltonians with open and periodic boundary conditions, respectively, on the lattice with side length $L=2^n$.
We assume that the two Hamiltonians satisfy the bounds\footnote{
For any hermitian matrices $\hA$ and $\hB$, we write $\hA\le\hB$ if and only if $\bph\hA\kph\le\bph\hB\kph$ for any $\kph$.
}
\eq
-b\,L^{d-1}\le \hHp_n-\hH_n\le b\,L^{d-1},
\lb{HpH}
\en
for any $n=1,2,\ldots$ with a constant $b>0$.
These are valid if the interactions are short-ranged.

\begin{figure}
\centerline{\includegraphics[width=10cm,clip]{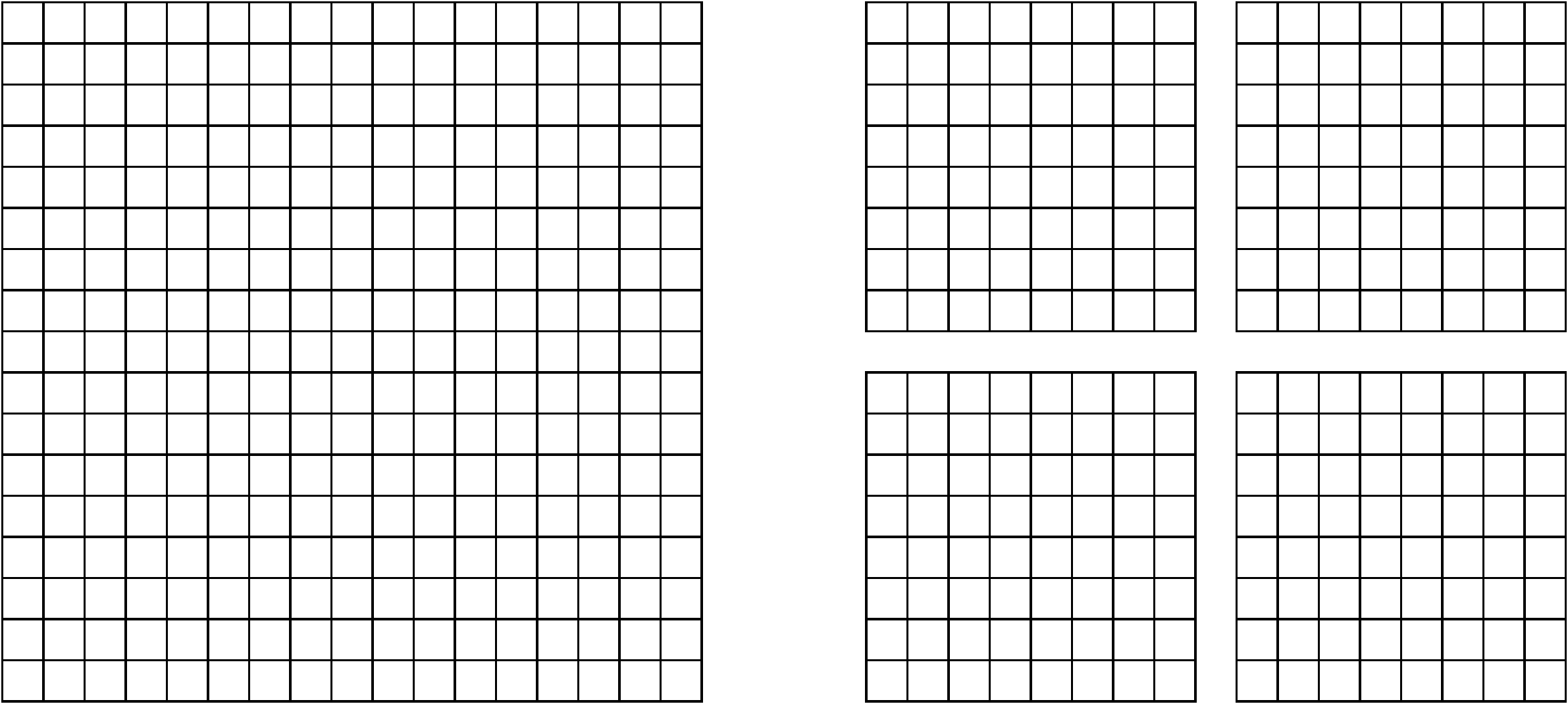}}
\caption[dummy]{
Decomposing the square lattice with $L=16$ into four copies of lattices with $L=8$.
}
\label{f:decomp}
\end{figure}

The lattice with side length $2^n$ can be naturally decomposed into $2^d$ copies of identical hypercubic lattice with side length $2^{n-1}$.
See Figure~\ref{f:decomp}.
We denote by $\hH_{n-1}^{(\nu)}$ with $\nu=1,\ldots,2^d$ the Hamiltonians (with open boundary conditions) on these sublattices.
Each  $\hH_{n-1}^{(\nu)}$ is identical to $\hH_{n-1}$.
The difference
\eq
\Di\hH_n=\hH_n-\sum_{\nu=1}^{2^d}\hH_{n-1}^{(\nu)}
\lb{dHn}
\en
describes the interaction between different sublattices.
We assume that there is a constant $c>0$, and
\eq
-c\,L^{d-1}\le\Di\hH_n\le0,
\lb{cLd}
\en
for any $n=1,2,\ldots$.
The lower bound clearly holds whenever the interactions are short-ranged.
The upper bound, on the other hand, is not valid in general, but can always be satisfied by subtracting suitable constants from the local Hamiltonian.\footnote{
For example, we replace the standard Heisenberg interaction $\hat{\boldsymbol{S}}_x\cdot\hat{\boldsymbol{S}}_x$ by $\hat{\boldsymbol{S}}_x\cdot\hat{\boldsymbol{S}}_x-S^2$.
}
This seemingly meaningless convention makes some book keeping easier.
See \rlb{OO0}.

\subsubsection{Useful Lemma}
Let $\hX$ be an arbitrary  self-adjoint operator.
For any $x\in\bbR$, we denote by $\Omega[\hX\le x]$ the number of eigenstates of $\hX$ with eigenvalues not larger than $x$.

\bigskip\noindent
{\em Lemma}\/: Let $\hX$ and $\hY$ be self-adjoint operators which satisfy
\eq
p\le\hX-\hY\le q,
\lb{aXYb}
\en
for some $p,q\in\bbR$.
Let $x_{\rm min}, x_{\rm max}, y_{\rm min}$, and $y_{\rm max}$ be the minimum and the maximum eigenvalues of $\hX$ and $\hY$.
Then one has
\eqg
p\le x_{\rm min}-y_{\rm min}\le q,\quad
p\le x_{\rm max}-y_{\rm max}\le q,
\lb{pq}\\
e^p\Tr[e^{\hY}]\le\Tr[e^{\hX}]\le e^q\Tr[e^{\hY}],
\lb{eab}\\
\Omega[\hY\le r-p]\ge\Omega[\hX\le r]\ge\Omega[\hY\le r-q],
\lb{Oca}
\eng
for any $r\in\bbR$.

\bigskip\noindent
{\em Proof}\/: Let $x_j$ and $y_j$ be eigenvalues of $\hX$ and $\hY$, respectively, which are ordered as $x_j\le x_{j+1}$ and  $y_j\le y_{j+1}$.
It is the standard implication of the mini-max principle (see, e.g., Corollary~III.1.2 of \cite{Bhatia}) that the assumption \rlb{aXYb} implies $y_j+p\le x_j\le y_j+q$ for each $j$.
Then all the claims are readily verified.\footnote{
The bounds \rlb{pq} can be proved using the elementary variational principle.
}~\qedm

\subsubsection{The minimum and the maximum energy densities}
\label{a:e}
We start by showing the existence of the limits in  \rlb{uminmax}, which define $\umin$ and $\umax$.
This is quite elementary.

Let $E^{\rm min}_n$ be the minimum eigenvalue (i.e., the ground state energy) of $\hH_n$.
The bounds \rlb{cLd}, along with \rlb{pq}, imply $-c\,2^{n(d-1)}\le E^{\rm min}_n-2^d E^{\rm min}_{n-1}\le0$.
This means that the density $\tiu^{\rm min}_n:=E^{\rm min}_n/2^{nd}$ satisfies $-c\,2^{-n}\le \tiu^{\rm min}_n-\tiu^{\rm min}_{n-1}\le0$.
The limit $\limn\tiu^{\rm min}_n$ exists because $(\tiu^{\rm min}_n)_{n=1,2,\ldots}$ is a Cauchy sequence.

If we denote by $E^{\rm per,min}_n$ the maximum and the minimum eigenvalue of $\hHp_n$, the bounds \rlb{HpH} with \rlb{pq} readily imply that $|E^{\rm per,min}_n/2^{nd}-u^{\rm min}_n|\le b\,2^{-n}$.
We conclude that the limit in \rlb{uminmax} for $\umin$ exists (for $N=2^{nd}$ with $n=1,2,\ldots$).
The proof for $\umax$ is exactly the same.

\subsubsection{Free energy}
\label{a:f}
The existence of the limit \rlb{philim}  of the free energy (the Massieu function) is proved in a similar and elementary manner.
For any $\beta\in(0,\infty)$, we find from \rlb{cLd} and \rlb{eab} that
\eq
\bigl\{\Tr[e^{-\beta\hH_{n-1}}]\bigr\}^{2^d}
\le\Tr[e^{-\beta\hH_n}]\le e^{\beta c L^{d-1}}\bigl\{\Tr[e^{-\beta\hH_{n-1}}]\bigr\}^{2^d},
\lb{TrTrTre}
\en
where we noted that\footnote{
In the expression  $\Tr[e^{-\beta\hH_{n-1}}]$, the trace is over the Hilbert space corresponding to the lattice with side length $2^{n-1}$.
} $\Tr[e^{-\beta\sum_{\nu=1}^{2^d}\hH_{n-1}^{(\nu)}}]=\{\Tr[e^{-\beta\hH_{n-1}}]\}^{2^d}$.
Define $\tiphi_n(\beta)=2^{-nd}\log\Tr[e^{-\beta\hH_n}]$, which is nothing but the Massieu function for free boundary conditions.
From \rlb{TrTrTre}, we find $0\le\tiphi_n(\beta)-\tiphi_{n-1}(\beta)\le c\beta2^{-n}$, which means that $(\tiphi_n(\beta))_{n=1,2,\ldots}$, for a fixed $\beta$, is a Cauchy sequence and hence the limit $\limn\tiphi_n(\beta)$ exists.
It is also clear that the convergence is uniform in any closed interval.

Noting that \rlb{HpH} implies $|\varphi_{2^{nd}}(\beta)-\tiphi_n(\beta)|\le b\beta\,2^{-n}$, we find that the desired limit \rlb{philim} exists  (for $N=2^{nd}$ with $n=1,2,\ldots$).
From the definition \rlb{massieu}, we see that $\varphi_N(\beta)$ is convex in $\beta$ for each $N$.
The limit $\varphi(\beta)$ is automatically convex.

\subsubsection{Entropy}
\label{a:s}
The infinite volume limit \rlb{sigma2} of the entropy density is much harder to control.

From \rlb{cLd} and \rlb{Oca}, we find for any $U$ that
\eq
\Omega[\hH_n\le U]\ge\Omega\Bigl[\,\sum_{\nu=1}^{2^d}\hH_{n-1}^{(\nu)}\le U\Bigr].
\lb{OO0}
\en
Noting that the Hamiltonians $\hH_{n-1}^{(1)},\dots,\hH_{n-1}^{(2^d)}$ can be diagonalized simultaneously, we see that the right-hand side can be bounded as\footnote{
Of course $\Omega[\hH_{n-1}\le U_\nu]$ is defined by regarding $\hH_{n-1}$ as an operator on the Hilbert space corresponding to the lattice with side length $2^{n-1}$.
}
\eq
\Omega\Bigl[\,\sum_{\nu=1}^{2^d}\hH_{n-1}^{(\nu)}\le U\Bigr]\ge\prod_{\nu=1}^{2^d}\Omega[\hH_{n-1}\le U_\nu],
\lb{OO1}
\en
for any $U_1,\ldots,U_{2^d}$ such that $\sum_{\nu=1}^{2^d}U_\nu=U$.
By choosing $U_\nu=U/2^d$ for all $\nu$, we see from \rlb{OO0} and \rlb{OO1} that
\eq
\Omega[\hH_n\le U]\ge\bigl\{\Omega[\hH_{n-1}\le U/2^d]\bigr\}^{2^d}.
\lb{OO2}
\en
We thus find that the entropy density $\tisigma_n(u)=2^{-nd}\log\Omega[\hH_n\le 2^{nd}u]$ for free boundary conditions satisfies 
\eq
\tisigma_n(u)\ge\tisigma_{n-1}(u)
\en
for each $u$ and $n=1,2,\ldots$.
Since we obviously have $\tisigma_n(u)\le\log(2S+1)$, we conclude that the nondecreasing series $(\tisigma_n(u))_{n=1,2,\ldots}$ is convergent for each $u$.
We define
\eq
\sigma(u):=\limn\tisigma_n(u).
\lb{sigmalim}
\en
Note that we have only proved the point-wise convergence.
Nevertheless we know that $\sigma(u)$ is nondecreasing in $u$ since each $\tisigma_n(u)$ is nondecreasing.
It is also obvious that $\sigma(u)\ge0$ for any $u\in(\umin,\umax)$.

We next repeat the decomposition as in Figure~\ref{f:decomp} $k$ times, and decompose the lattice with side length $2^n$ into $2^{kd}$ identical copies of lattices with side length $2^{n-k}$.
Exactly as in \rlb{OO0} and \rlb{OO1}, we obtain
\eq
\Omega[\hH_n\le U]\ge\prod_{\nu=1}^{2^{kd}}\Omega[\hH_{n-k}\le U_\nu],
\lb{OO3}
\en
for any $U_1,\ldots,U_{2^{kd}}$ such that $\sum_{\nu=1}^{2^{kd}}U_\nu=U$.
Take any $u_1,u_2\in(\umin,\umax)$ and $q=1,2,\ldots,2^{kd}-1$, and set 
$U_\nu=2^{(n-k)d}u_1$ for $\nu=1,\ldots,q$ and $U_\nu=2^{(n-k)d}u_2$ for $\nu=q+1,\ldots,2^{kd}$.
Then \rlb{OO3} implies
\eq
\Omega[\hH_n\le 2^n\{\lambda u_1+(1-\lambda)u_2\}]
\ge
\bigl(\Omega[\hH_{n-k}\le2^{(n-k)d}u_1]\bigr)^q
\bigl(\Omega[\hH_{n-k}\le2^{(n-k)d}u_2]\bigr)^{2^{kd}-q},
\en
where $\lambda=q/2^{kd}$.
This implies
\eq
\tisigma_n(\lambda u_1+(1-\lambda)u_2)\ge\lambda\,\tisigma_{n-k}(u_1)+(1-\lambda)\tisigma_{n-k}(u_2).
\en
By letting $n\up\infty$, we find that
\eq
\sigma(\lambda u_1+(1-\lambda)u_2)\ge\lambda\,\sigma(u_1)+(1-\lambda)\sigma(u_2),
\lb{convpre}
\en
for any $u_1,u_2\in(\umin,\umax)$ and $\lambda\in(0,1)$ which is written as $\lambda=q/2^{kd}$ for some $k=1,2,\ldots$ and $q=1,2,\ldots,2^{kd}-1$.
This almost looks like the definition of concavity, but $\lambda$ is limited to special values.
Slightly more work is needed.

Take arbitrary $u_-$ and $u_0$ such that $\umin<u_-<u_0<\umax$, and fix them.
Take any $\lambda=q/2^{kd}$ (with $k=1,\ldots$ and $q=1,\ldots,2^{kd}-1$) such that
\eq
\delta=\frac{\lambda}{1-\lambda}(u_0-u_-)
\lb{delta}
\en
satisfies $u_0+\delta\le\umax$.
Noting that $\lambda u_-+(1-\lambda)(u_0+\delta)=u_0$, we get from \rlb{convpre} that
\eq
\sigma(u_0)\ge\lambda\sigma(u_-)+(1-\lambda)\sigma(u_0+\delta),
\en
which implies
\eqa
\sigma(u_0+\delta)&\le\frac{1}{1-\lambda}\sigma(u_0)-\frac{\lambda}{1-\lambda}\sigma(u_-)
\nl&=\sigma(u_0)+\frac{\sigma(u_0)-\sigma(u_-)}{u_0-u_-}\,\delta.
\ena
Since we have $\sigma(u_0+\delta)\ge\sigma(u_0)$, we find that $\sigma(u_0+\delta)$ approaches $\sigma(u_0)$ when $\delta$ approaches zero through the values of the form \rlb{delta}.
Also noting that $\sigma(u_0+\delta)$ is non-decreasing in $\delta>0$, this implies that $\sigma(u_0+\delta)\to\sigma(u_0)$ as $\delta\downarrow0$ (where $\delta$ takes any real values), i.e, $\delta(u)$ is right-continuous at $u_0$.
The left continuity is proved in exactly the same manner.

We thus conclude that $\sigma(u)$ is continuous over the whole interval $(\umin,\umax)$.
Then  \rlb{convpre} is enough to prove that $\sigma(u)$ is concave.
Finally since $\tisigma_n(u)$ is nondecreasing in $n$ and $\sigma(u)$ is continuous, Dini's theorem ensures that the convergence is uniform in any closed interval.

It remains to treat the entropy density of models with periodic boundary conditions.
Note that \rlb{HpH} and \rlb{Oca} imply
\eq
\Omega[\hH_n\le U+c\,2^{n(d-1)}]\ge\Omega_{2^{nd}}(U)\ge\Omega[\hH_n\le U-c\,2^{n(d-1)}],
\en
for any $u\in(\umin,\umax)$, where $\Omega_N(U)$ is the number of states for periodic boundary conditions (as in the main text).
By setting $U=2^{nd}u$, we get
\eq
\tisigma_n(u+c\,2^{-n})\ge\sigma_{2^{nd}}(u)\ge\tisigma_n(u-c\,2^{-n}).
\en
Both the left-hand and right-hand sides converge to $\sigma(u)$ as $n\up\infty$ because of the uniform convergence, and so does $\sigma_{2^{nd}}(u)$.
The uniformity of the convergence is obvious.

\subsection{Equivalence of ensembles}
\label{s:sigmaD}
\paragraph{Standard thermodynamic functions}
The proof of the equivalence of ensembles for thermodynamic functions is standard and elementary.
Let $\beta_0\in(0,\infty)$.
For any $U$, we see from the definition \rlb{ZN} that
\eq
Z_N(\beta_0)\ge\mathop{\sum_{j=1}^{\GN}}_{(E_j\le U)}e^{-\beta_0 \EN_j}
\ge \mathop{\sum_{j=1}^{\GN}}_{(E_j\le U)}e^{-\beta_0 U}=\Omega_N(U)\,e^{-\beta_0 U},
\en
which in particular implies
\eq
Z_N(\beta_0)\ge\max_U\Omega_N(U)\,e^{-\beta_0 U}.
\lb{ZN>}
\en

Denote by $\tiU$ the value of $U$ which maximizes the right-hand side of \rlb{ZN>} (with $\beta_0$ and $N$ fixed).
As in \rlb{tdbound}, take a sequence $\tiU_0,\tiU_1,\ldots,\tiU_M$ such that (i)~$\tiU_\nu=\tiU$ for one $\nu$, (ii)~$\tiU_{\nu}-\tiU_{\nu-1}=\delta$ for $\nu=1,\ldots,M$, and (iii)~$(\tiU_0,\tiU_M)\ni E_j$ for all $j=1,\ldots,\GN$.
Then we have
\eqa
Z_N(\beta_0)&\le\sum_{\nu=1}^{M}D_N(\tiU_\nu,\delta)\,e^{-\beta_0\tiU_{\nu-1}}
\le e^{\beta_0\delta}\sum_{\nu=1}^{M}\Omega_N(\tiU_\nu)\,e^{-\beta_0\tiU_{\nu}}
\nl
&\le M\,e^{\beta_0\delta}\,\Omega_N(\tiU)\,e^{-\beta_0\tiU}
=M\,e^{\beta_0\delta}\max_U\Omega_N(U)\,e^{-\beta_0 U}.
\lb{ZN<}
\ena

By noting that $M\propto N$, we find from \rlb{ZN>} and \rlb{ZN<} that
\eq
\limN\frac{1}{N}\log Z_N(\beta_0)=\limN\max_U\frac{1}{N}\{\log\Omega_N(U)-\beta_0U\},
\lb{ZN=}
\en
which immediately implies
\eq
\varphi(\beta_0)=\limN\max_u\{\sigma_N(u)-\beta_0u\}=\max_u\{\sigma(u)-\beta_0u\},
\en
where we exchanged the limit and the max, noting that the convergence is  uniform.
We have thus proved \rlb{Legendre2}.
The other relation \rlb{Legendre} follows from the general theory of the Legendre transformation.

\paragraph{Different expression of entropy}
We assume that $\varphi(\beta)$ is differentiable at $\beta=\beta_0\in(0,\infty)$, and let $u_0=-\varphi'(\beta_0)$.
This excludes the possibility of phase coexistence at $u_0$.

Take energy width $\Delta_N$ such that $\Delta_N\ge\delta$ for any $N$, where $\delta>0$ is an arbitrary fixed constant.
We define
\eq
U_N(\beta_0,\Delta_N):=\mathop{\text{arg-max}}_{U}\{\log D_N(U,\Delta_N)-\beta_0U\}.
\lb{UN2}
\en
Note that $U_N(\beta_0)$ in the main text is $U_N(\beta_0,\delta)$ in this notation.

We then show the following.

\bigskip\noindent
{\em Proposition}\/:
Under the above assumption, we have
\eqg
\limN\frac{1}{N}U_N(\beta_0,\Delta_N)=u_0,
\lb{UND}\\ 
\limN\frac{1}{N}\log D_N(U_N(\beta_0,\Delta_N),\Delta_N)=\sigma(u_0).
\lb{DUND}
\eng

\bigskip

The relation \rlb{DUND} with \rlb{UND} is similar to the relation \rlb{sigma3}, which is conjectured to be valid whenever $\Delta_N\up\infty$ as $N\up\infty$ (but proved only when $\Delta_N\ge N\Di u$).
But the two relations are different since \rlb{DUND} makes use of the ``fine-tuned'' energy range determined by \rlb{UN2}.

\bigskip\noindent
{\em Proof}\/:
Repeating the derivation of \rlb{ZN=}, one can prove
\eqa
\limN\frac{1}{N}\log Z_N(\beta_0)&=\limN\frac{1}{N}\max_U\bigl\{\log D_N(U,\Delta_N)-\beta_0 U\bigr\}
\nl
&=\limN\frac{1}{N}\bigl\{\log D_N(U_N(\beta_0,\Delta_N),\Delta_N)
-\beta_0\,U_N(\beta_0,\Delta_N)\bigr\},
\lb{DU1}
\ena
where the final expression follows from the definition of $U_N(\beta_0,\Delta_N)$.

Since $U_N(\beta_0,\Delta_N)/N$ is bounded one can take a subsequence of $N$ in which $U_N(\beta_0,\Delta_N)/N$ converges to $u^*$.
Noting that $\Delta_N(U,\Delta_N)\le\Omega_N(U)$ and using  \rlb{sigma2}, we find
\eq
\lim_{N\up\infty}\frac{1}{N}\bigl\{\log D_N(U_N(\beta_0,\Delta_N),\Delta_N)
-\beta_0\,U_N(\beta_0,\Delta_N)\bigr\}\le\sigma(u^*)-\beta_0u^*,
\lb{DU2}
\en
where $N$ in the left-hand side is taken from the subsequence.
But since \rlb{Legendre2} implies 
\eq
\limN\frac{1}{N}\log Z_N(\beta_0)=\max_u\{\sigma(u)-\beta_0u\},
\en
and the maximum is attained only at $u=u_0$, \rlb{DU1} and \rlb{DU2} imply $u^*=u_0$.
This proves \rlb{UND}.
Then \rlb{DU1} implies \rlb{DUND}.~\qedm

\bigskip
{\small
I wish to thank Takahiro Sagawa for suggesting the problem and for valuable discussions and comments.
I also thank 
Yoshiko Ogata,
Tohru Koma,
Taku Matsui,
and 
Daniel Ueltschi for useful discussions and comments.
The present work was supported by JSPS Grants-in-Aid for Scientific Research no.~16H02211.
}


\end{document}